# Memory-Efficient Super-Resolution of 3D Micro-CT Images Using Octree-Based GANs: Enhancing Resolution and Segmentation Accuracy

Evgeny Ugolkov[a], Xupeng He[b], Hyung Kwak[b], Hussein Hoteit[a]

[a] Physical Science and Engineering, King Abdullah University of Science and Technology (KAUST), Saudi Arabia

[b] EXPEC Advanced Research Center, Aramco, Dhahran, Saudi Arabia

*Corresponding author: Hussein.hoteit@kaust.edu.sa

**Abstract**

We present a memory-efficient algorithm for significantly enhancing the quality of segmented 3D micro-Computed Tomography (micro-CT) images of rocks using a Machine Learning (ML) Generative Model. The proposed model achieves a 16x increase in resolution and corrects inaccuracies in segmentation caused by the overlapping X-ray attenuation in micro-CT measurements across different minerals. The generative model employed is a 3D Octree-Based Progressive Growing Deep Convolutional Wasserstein Generative Adversarial Network with Gradient Penalty (3D OB PG DC WGAN-GP). To address the challenge of extremely high memory consumption inherent in standard PyTorch 3D Convolutional (Conv3D) layers, which is a significant constraint in 3D Super-Resolution (SR) applications, we implemented an Octree structure within the 3D Progressive Growing Generator (3D PG G) model. This enabled the use of memory-efficient 3D Octree-Based Convolutional layers provided by the open-source Minkowski Engine library. The adoption of the octree structure was pivotal in overcoming the long-standing memory bottleneck in volumetric deep learning, making it possible to reach 16× Super-Resolution in 3D, a scale that is challenging to attain due to cubic memory scaling. For training, we utilized segmented 3D Low-Resolution (LR) micro-CT images along with unpaired segmented complementary 2D High-Resolution (HR) Laser Scanning Microscope (LSM) images. Post-training, we achieved high-quality, segmented 3D SR images with resolutions improved from 7 to 0.44 µm/voxel and accurate segmentation of constituent minerals. Validated on Berea sandstone, this framework demonstrates substantial improvements in pore characterization and mineral differentiation, which are key factors for accurate Digital Rock Physics (DRP) simulations. The proposed algorithm advances the feasibility of large-scale, high-resolution 3D reconstructions and offers a robust solution to one of the primary computational limitations in modern geoscientific imaging.

## 1. Introduction

Digital Rock Physics (DRP) plays a critical role in accurately characterizing the pore-scale structure and mineral composition of rocks, influencing our understanding of subsurface rock formations and guiding decisions across various sectors, including energy production and environmental management. However, conventional micro-Computed Tomography (micro-CT) imaging methods often lack sufficient resolution to capture crucial sub-micron features, resulting in inaccurate or incomplete digital representations.

Enhancing the resolution of micro-CT images is therefore critical for improving the reliability of DRP simulations and subsequent analyses. Recent advancements in machine learning, particularly Generative Adversarial Networks (GANs), offer promising solutions for image Super-Resolution tasks. Previously developed methods have demonstrated notable improvements in resolution and segmentation accuracy, enabling the visualization of sub-micron features beyond conventional imaging capabilities. Ugolkov et al. (2025) introduced a Super-Resolution algorithm for segmented 3D micro-CT images, achieving an 8x scale factor and refining the voxel size from 3.5 µm to 0.4375 µm. The algorithm successfully added sub-micron features and corrected inaccuracies in the segmentation process. The training was conducted using unpaired segmented 3D micro-CT and 2D LSM images, allowing for convenient and independent acquisition of 3D and 2D datasets. The High-Resolution dataset was manually segmented and significantly expanded using StyleGAN2ADA. The training utilized the WGAN-GP algorithm, with a 3D Generator and 2D Discriminator built with standard PyTorch 3D and 2D Convolutional (Conv3D and Conv2D) layers, respectively. The resulting 8x Super-Resolved 3D image maintained the Field of View (FOV) of a micro-CT image while achieving the resolution of an LSM image. However, significant computational challenges remain, as higher resolution (16x or higher) in 3D volumetric data typically demands exponential increases in memory usage and computational power. Consequently, our initial efforts to achieve a 16x scaling factor resulted in the training process terminating due to the Out-Of-Memory (OOM) error. This outcome was expected, given that micro-CT generates 3D voxelized images, and the use of standard PyTorch Conv3D layers across the entire 3D volume causes memory and computational costs to increase cubically with image size, which is a major limitation to achieve higher resolution.

This study addresses these computational constraints by introducing a memory-efficient Octree-Based Progressive Growing GAN architecture tailored specifically for 3D micro-CT image Super-Resolution.

Unlike standard convolutional neural networks that require substantial computational resources, the proposed octree-based approach significantly reduces memory demands, enabling higher resolution enhancements. Consequently, the method presented here successfully achieves a substantial 16x increase in resolution, effectively resolving sub-micron features critical for more accurate DRP simulations.

Based on the 3D shape representation, several classes of Convolutional Neural Network (CNN) methods have been presented for the 3D shape analysis (Wang et al., 2017). Manifold-based methods (Boscaini et al., 2016; Bronstein et al., 2016; Sinha et al., 2016) apply convolutional neural networks directly to geometric features defined on the surface of 3D mesh representations. Multiview-based methods (Bai et al., 2016; Qi et al., 2017) represent a 3D shape through a series of images rendered from various viewpoints, which are subsequently analyzed using 2D CNNs. However, our research focuses on voxel-based methods, as micro-CT imaging inherently produces voxelized 3D data.

Voxel-based methods represent 3D shapes as discrete functions defined on voxel grids and utilize 3D CNNs to analyze the entire volumetric data. Various strategies have been proposed to address the substantial memory requirements associated with such CNN methods. For instance, Wu et al., (2015) proposed 3D ShapeNets, aimed at object recognition and shape completion tasks, though memory and computational constraints restricted their application to relatively small voxel grids. Maturana and Scherer (2015) improved this approach by reducing the number of parameters per voxel. However, these full-voxel-based methods were limited to low resolutions of size $30^3$ voxels due to the high memory and computational cost.

To further optimize computational efficiency, Graham (2015) introduced 3D sparse CNNs, which perform CNN operations only on active voxels and activate adjacent voxels within the convolution kernel. However, as the number of convolution layers between pooling layers increases, the efficiency of this method diminishes. An alternative, Octree-based CNNs (Wang et al., 2017) presented a method, named O-CNN, to represent 3D shapes with octrees and perform 3D CNN operations only on the sparse octants occupied by the boundary surfaces of 3D shapes. Although efficient for certain applications, these octree methods are not ideal for micro-CT rock images due to their dense, highly irregular, intertwined structures with multi-mineral, multi-phase composition and micron-size components. This makes them unsuitable for representation as boundary surfaces. Another octree-based method, OctNet (Riegler et al., 2016), assumed a predefined octree structure known at inference, making it unsuitable for super-resolution tasks where the structure is not known in advance and must be predicted.

To address these limitations, Tatarchenko et al. (2017) introduced the Octree-Generating Network (OGN), which progressively refines octrees during decoding, significantly reducing memory consumption while maintaining high-resolution detail. This concept provides an effective basis for our memory-optimized super-resolution approach. Starting from a specific layer in the network, dense regular grids were replaced by octrees. Therefore, the OGN predicts large uniform regions in the output space during the early decoding stages, reducing the computational load for the subsequent high-resolution layers. Only areas with intricate details are processed by the more resource-intensive layers. As a result, OGN achieves the same level of accuracy as traditional dense decoders while using considerably less memory and operating much faster at high resolutions.

This concept served as the foundational inspiration for the memory-efficient optimization strategy adopted in our work. While the original OGN implementation was developed using C++ and the Caffe deep learning framework, the authors noted that the associated GitHub repository was no longer actively maintained. However, they referenced the Minkowski Engine (Choy et al., 2019), an open-source library that offers similar functionality with support for sparse convolutional operations in PyTorch. Building upon these advancements, we developed our Octree-Based implementation for the 3D Generator model using the Minkowski Engine, enabling efficient super-resolution of volumetric micro-CT data with drastically reduced memory requirements.

Building upon these advances and addressing the computational challenges outlined above, we introduce a novel memory-efficient algorithm: a 3D Octree-Based Progressive Growing Deep Convolutional Wasserstein Generative Adversarial Network with Gradient Penalty (3D OB PG DC WGAN-GP). Our approach specifically targets super-resolution of 3D micro-CT images of rocks, significantly improving their resolution and rectifying segmentation inaccuracies in both pore spaces and mineral phases. The model uses 3D segmented Low-Resolution micro-CT images of rocks with unpaired 2D segmented High-Resolution Laser Scanning Microscope (LSM) images made on thin sections. Our model efficiently refines voxel resolution, thereby capturing crucial sub-micron structural and mineralogical features previously unresolved in standard micro-CT imaging. Furthermore, the images of arbitrarily large sizes can be generated from the chunks of input 3D Low-Resolution images, allowing the capture of larger-scale heterogeneities. In the following sections, we detail our training dataset preparation, octree-based data structure integration, model architecture, training algorithm, and demonstrate the achieved improvements in resolution and segmentation accuracy.

The remainder of this paper is organized as follows: Section 2 details the materials and methods, including dataset preparation, octree structure integration, and the architecture of the proposed model.

Section 3 presents the results of applying the algorithm to Berea sandstone, along with quantitative and qualitative evaluations. Section 4 discusses the implications of the findings in the context of Digital Rock Physics. Finally, Section 5 summarizes the conclusions and outlines potential directions for future work.

## 2. Materials and methods

### 2.1. Segmented 3D Low-Resolution training dataset

A Berea sandstone plug with a diameter of 12 mm was prepared to obtain high-quality imaging using a Tescan CoreTOM X-ray micro-CT scanner. The raw 3D micro-CT image acquisition and reconstruction were performed with Acquila software. The micro-CT scan was performed at 80 kV and 15 W, achieving a voxel resolution of 7 μm, with the resulting image averaged over 20 acquisitions. To enable the random sub-volume extraction algorithm and reduce beam hardening artifacts, the reconstructed cylindrical volume was cropped into a parallelepiped with dimensions of 1192 × 1266 × 1540 voxels. **Figure 1a** presents a cubic region of the 3D volume alongside a 2D slice of the raw micro-CT dataset, where the grayscale intensity values range from 0 to 65535.

The segmentation of the raw 3D micro-CT dataset was performed using the Segmentation Toolbox in Avizo software (Thermo Fisher Scientific, 2022), employing the Interactive Thresholding method. Threshold values for grayscale intensities corresponding to the pore space and different mineral groups were manually determined (**Table 1**) based on visual inspection and prior knowledge of the mineral composition (Ugolkov et al., 2025). The dataset was segmented into four distinct categories: pores, clay, quartz, and feldspar. Due to their low abundance, heavy minerals were included within the quartz category rather than classified separately. A representative segmented portion of the 3D volume along with a corresponding 2D slice is shown in **Figure 1b**.

Features smaller than the voxel size were unresolved due to the inherent resolution limits of micro-CT imaging, causing an underestimation of pore space and resulting in a significant portion of porosity being undetected. For instance, after segmentation, the calculated porosity of the 3D low-resolution micro-CT dataset was 5%, whereas the Helium porosimetry experiment indicated a porosity of 21%. Additionally, due to similar X-ray attenuation properties, certain groups may be misclassified, particularly between pores and clays or between feldspars and quartz (Golab et al., 2013). As a result, the segmentation of feldspar minerals was not entirely accurate, while the clay group didn't include sub-micron porosity. In subsequent sections, we will demonstrate that the proposed algorithm effectively addresses these challenges.

**Table 1.** Grayscale intensity thresholds used for segmentation of micro-CT images into different mineral and pore groups.

| Gray-scale values | Group |
|---|---|
| 0-4200 | Pore |
| 4200-4800 | Clay |
| 4800-7000; 8000-65535 | Quartz |
| 7000-8000 | Feldspar |

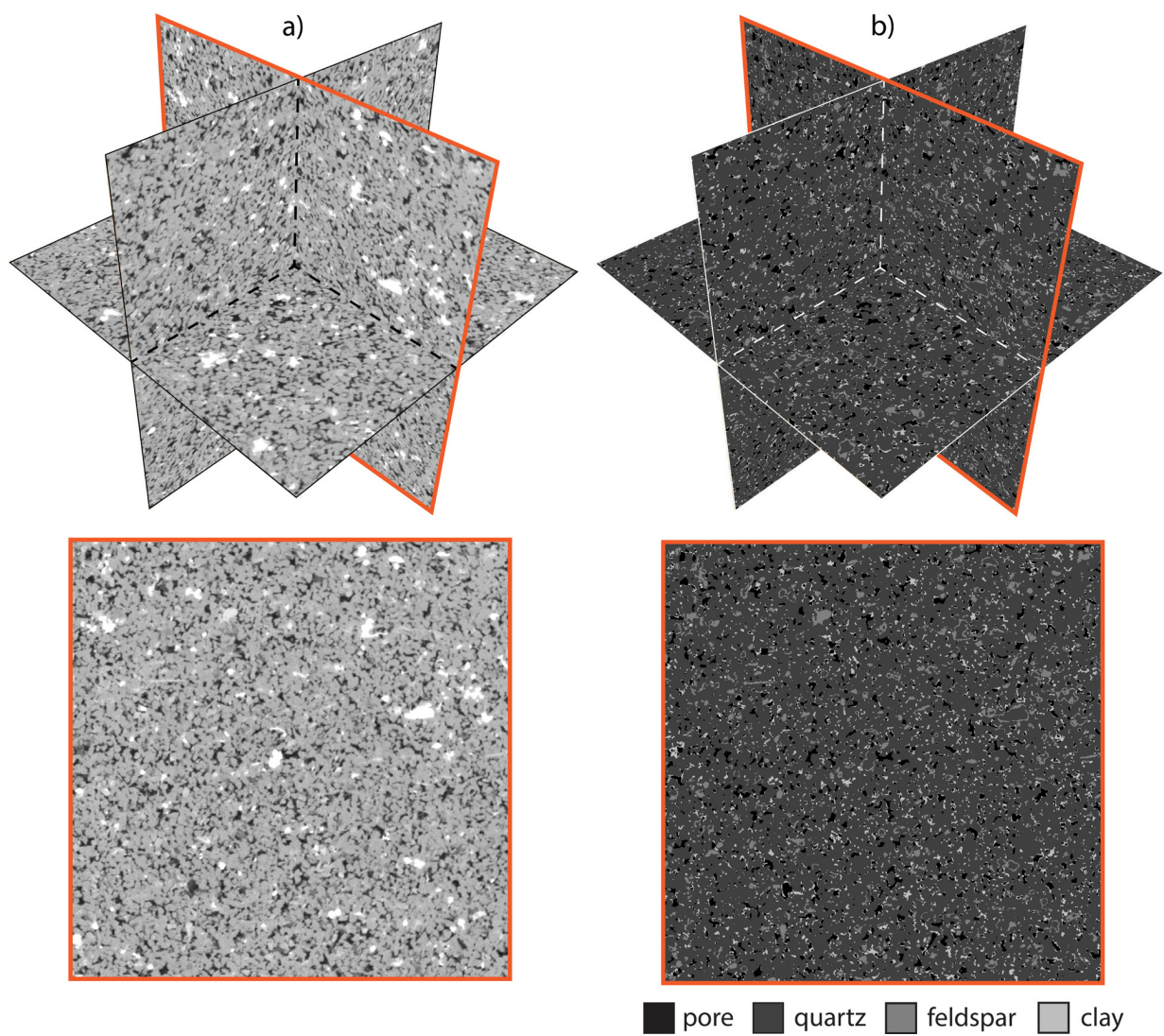

**Figure 1.** Comparison of raw (a) and segmented (b) micro-CT images illustrating the segmentation process used to differentiate pore space and mineral phases. The segmented image (b) highlights the distinct

groups: pores, clay, quartz, and feldspar. Images (b) correspond to the segmented 3D low-resolution training dataset.

### 2.2. Segmented 2D high-resolution training dataset

For the 2D high-resolution training dataset, we employed segmented 2D Laser Scanning Microscope (LSM) images consistent with those described in previous research (Ugolkov et al., 2025), which detailed the information regarding the acquisition, segmentation, and dataset enhancement procedures. Representative examples of the segmented 2D high-resolution LSM images utilized for training are illustrated in **Figure 2**. These segmented images distinctly classify the microstructure of Berea sandstone into four categories: pore space, quartz, feldspar, and clay, which highlight the mineralogical and pore space distribution. These images are used to enhance the training process of the proposed super-resolution model.

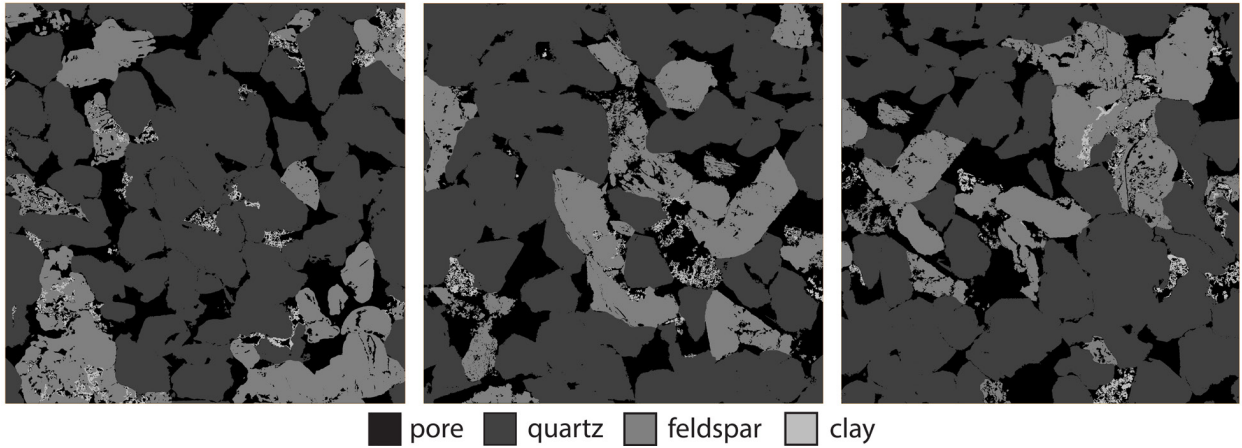

**Figure 2.** Examples of 2D high-resolution training images corresponding to segmented 2D LSM images of Berea sandstone, showing distinct mineralogical phases and pore spaces classified into four groups: pore space (black), quartz (dark gray), feldspar (medium gray), and clay (light gray).

### 2.3. Octree structure for micro-CT images of rocks

An octree is a hierarchical data structure commonly used to efficiently represent 3D data by recursively subdividing a cubic volume into eight smaller octants. This adaptive partitioning allows regions with complex, fine-scale details to be represented with higher resolution, while homogeneous regions are represented at coarser resolutions. In many learning tasks, neighboring voxels in a grid often share the same state, such as in a binary occupancy map or multi-class labeling of a 3D object or scene. In such cases, data can be efficiently represented using octrees.

The octree structure is commonly employed in computer graphics for tasks such as rendering, modeling, and collision detection. Introduced by (Meagher, 1982), an octree is a 3D grid with adaptive cell sizes, enabling a significant reduction in memory usage compared to a regular voxel grid without any loss of information. A function defined on a voxel grid can be transformed into one defined on an octree by starting with a single cell representing the entire space and recursively dividing cells into eight octants. If all voxels within a cell share the same function value, the cell remains undivided and becomes a leaf node in the tree. The collection of cells at a specific resolution is known as an octree level.

The choice for the octree structure was driven by the observation that only limited regions within the micro-CT images require improvements in resolution, particularly pore-grain boundaries, grain-grain contacts, and intragranular pores. These critical areas, referred to as "mixed" regions, exhibit substantial variability. Conversely, extensive areas such as the centers of mineral grains and pore spaces remain relatively uniform and unaffected by increased resolution; these we classify as "dense" regions. Consequently, 3D segmented micro-CT images inherently present significant opportunities for memory optimization, opportunities not leveraged by traditional PyTorch Conv3D implementations.

The octree data structure perfectly suits the segmentation configuration. With this method, a 3D space is recursively subdivided into eight smaller octants. Each node in an octree represents a cubic region of space. Each node is subdivided (or not subdivided) individually based on specific rules. In our super-resolution framework, the subdivision of a node depends on its content: nodes containing octants with differing groups (multiple phases or minerals) are subdivided further and labeled as "mixed," whereas nodes containing octants belonging exclusively to a single group (either pore or a single mineral) remain undivided and are labeled as "dense." This recursive subdivision process continues until predefined criteria, such as reaching the voxel size corresponding to the desired final resolution (the resolution of high-resolution LSM training images), are satisfied. **Figure 3** provides a schematic representation of this octree subdivision process for a simplified synthetic rock image with three distinct groups. For illustrative clarity, subdivision is depicted in a 2D context using quadrants, whereas our implemented algorithm applies this approach fully in 3D space using eight octants. This figure demonstrates the adaptive partitioning concept, highlighting how subdivision progressively refines spatial resolution around regions of complexity. Starting from Level 0, the entire region is initially evaluated as a single node. In subsequent levels, nodes identified as "mixed" (containing multiple dense groups) are recursively subdivided into four smaller quadrants. Nodes classified as "dense," (consisting entirely of a single group) remain undivided and are represented at coarser resolutions. This selective subdivision process continues incrementally from Level 0 through Level 5, refining detail only where necessary and significantly reducing computational requirements by avoiding unnecessary subdivisions in uniform regions. While this figure demonstrates the subdivision in a two-dimensional space for clarity, the implemented algorithm operates fully in 3D, subdividing nodes into eight octants rather than quadrants.

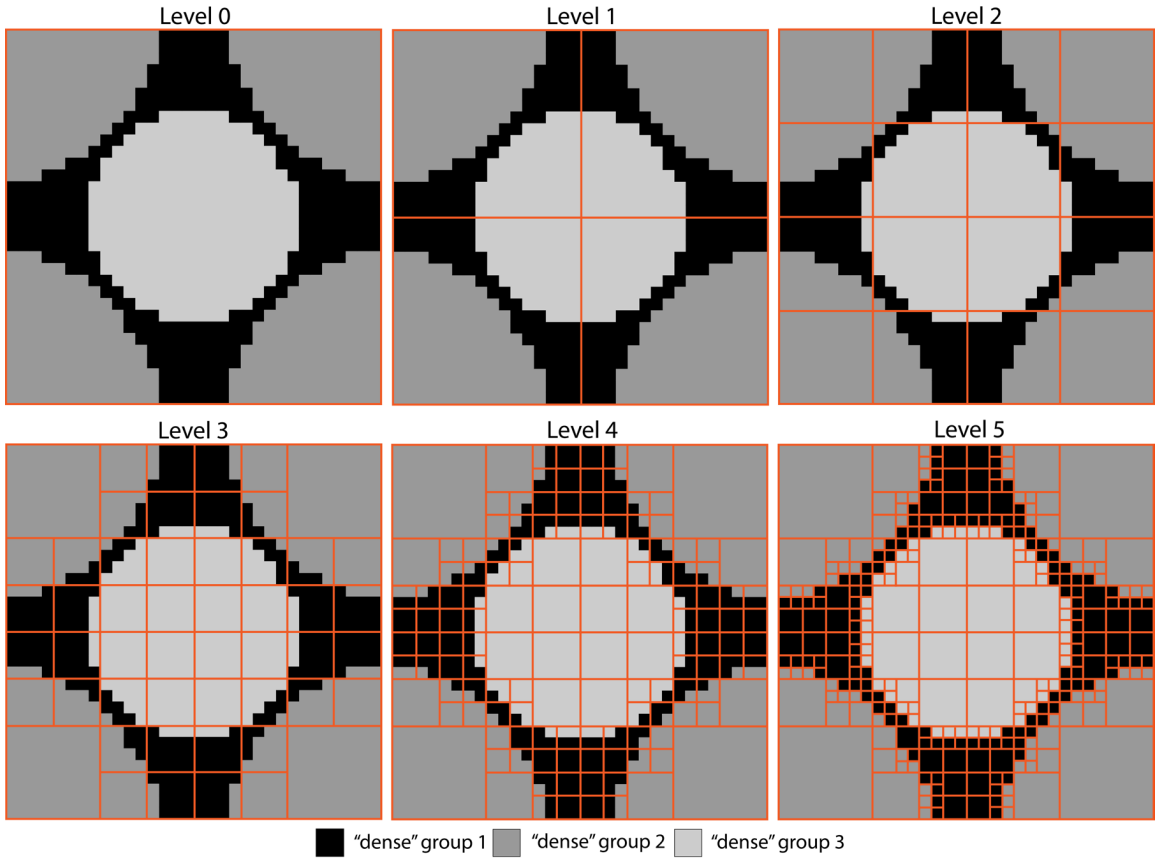

**Figure 3.** Schematic illustration of the octree subdivision process using a simplified synthetic rock image containing three distinct dense groups. Each level demonstrates adaptive partitioning, recursively subdividing only "mixed" nodes (regions containing multiple groups) while preserving "dense" nodes (regions of uniform composition) at coarser resolutions. This figure is shown in 2D for clarity, whereas the actual implementation applies octree subdivision fully in 3D.

### 2.4. Octree-Based Progressive Growing Generator

Integrated into our super-resolution training framework, the octree structure enables selective processing of "mixed" nodes while bypassing computations in "dense" nodes that contain just a single group. This approach efficiently consolidates substantial clusters of spatially adjacent voxels with identical values into coarse-resolution cells. As a result, computational resources are reallocated to nodes containing fine-scale details, enhancing efficiency without compromising resolution fidelity.

To enable our model to distinguish between "mixed" and "dense" nodes, we employed the Progressive Growing (PG) algorithm, originally introduced by (Karras et al., 2017) for 2D image generation. The PG approach structures training into sequential "Stages," wherein both the model complexity and the processed volume progressively expand across successive Stages. In our Octree-Based (OB) adaptation, however, only "mixed" nodes propagate to the subsequent Stages, while "dense" nodes are stored and excluded from further computations.

Conceptually, in the PG framework, the training process follows a structured sequence of classification, pruning, and subdivision (**Figure 4**). At each resolution stage, the classification layer assigns each node in the input volume a unique group: a specific mineral, a pore, or a "mixed". The "mixed" classification indicates that, at the current resolution, the node cannot be definitively attributed to a single phase, necessitating further refinement through subdivision. The pruning mechanism ensures that only "mixed" nodes proceed to the next resolution stage, while "dense" nodes, those entirely belonging to a single group, are excluded from further computations and stored for future reconstructions. In the subsequent resolution stage, each of the proceeded "mixed" nodes is subdivided with the ME Block, which is a sequence of Transpose Convolution, Batch Normalization, and Convolutional layers (**Table A1**), into 8 smaller octants, and the classification process is repeated at this finer scale. If a newly generated octant is classified as a single mineral or pore, it is stored and removed from further processing. Conversely, if an octant is classified as "mixed," it is passed to the next stage for further subdivision. This iterative classification-pruning-subdivision cycle continues until the target high-resolution training image resolution is achieved.

The 3D Progressive Growing Generator (3D PG G) leverages octree-based data structures for efficient 3D processing. We implemented the 3D PG G model utilizing the Minkowski Engine (Choy et al., 2019), an open-source framework that enables efficient Octree-Based Conv3D operations. This approach significantly reduces memory requirements compared to traditional dense PyTorch Conv3D implementations. Consequently, our method can handle substantially larger model sizes before encountering Out-of-Memory (OOM) errors, enabling higher-quality super-resolution results at larger scales (**Figure 5**).

The proposed algorithm is a general multi-stage progressive growing framework designed for 3D super-resolution tasks. It can be adapted to different scaling factors based on resolution requirements. In this implementation, we apply it for a 16x super-resolution scale factor, including five stages where each stage progressively refines the resolution by a factor of two, ensuring efficient memory utilization and accurate reconstruction of high-resolution 3D micro-CT images. The main functionalities of the stages are defined as:

**Stage 1** (coarse feature extraction): The process begins with this stage, where a randomly sampled $32^3$ input volume at an initial resolution of 7 μm/voxel is analyzed. At this stage, the model predicts and classifies regions into two categories: "mixed" (requiring further refinement) and "dense" (homogeneous and finalized). The identified dense regions are stored and removed from further computations, while only mixed nodes are forwarded to the next stage for further refinement.

**Stages 2-4** (hierarchical refinement and subdivision): In these stages, the model continues refining the remaining *mixed* nodes. Each stage subdivides the received nodes into eight octants, effectively doubling the resolution in each dimension. The newly created octants are then classified, determining which of them are *dense* and which remain *mixed*. The dense octants are stored, while the *mixed* octants continue propagating through the subsequent stages for further subdivision and refinement. This hierarchical refinement ensures that only the most complex regions undergo additional processing, significantly reducing computational costs while preserving fine-scale details.

**Stage 5** (final subdivision and classification): This is the final step in the process, where the remaining *mixed* nodes undergo a last round of subdivision and classification. At this stage, the model produces a fully segmented 3D super-resolution image with a final resolution of 0.4375 μm/voxel, achieving a 16x enhancement over the input image.

Additionally, Stages 1-4 reconstruct intermediate dense segmented 3D Super-Resolution images with corresponding resolution using the newly predicted nodes and all the memorized on the prior Stages "dense" nodes. The newly predicted nodes remain in the current resolution, while memorized "dense" nodes are subdivided multiple times, depending on the Stage number they were stored while keeping the same value for each new octant to achieve the current resolution level. These intermediate dense segmented 3D Super-Resolution images are the foundation for the Progressive Growing training algorithm, as described in Section 2.7.

During training, both the 3D Progressive Growing Generator (3D PG G) and 2D Progressive-Growing Discriminator (2D PG D) (Section 2.6.) models build new layers on top of the pre-trained layers from the previous Stage. Both 3D PG G and 2D PG D models are trainable from the beginning to the end and don't contain non-differentiable operations, guaranteeing smooth gradient flow. **Figure 4** schematically demonstrates the 3D Progressive Growing Generator workflow. The architecture of the 3D PG G with colors matching **Figure 4** is presented in **Table A1**. The algorithm for the 3D OB PG G is presented in **Appendix D**.

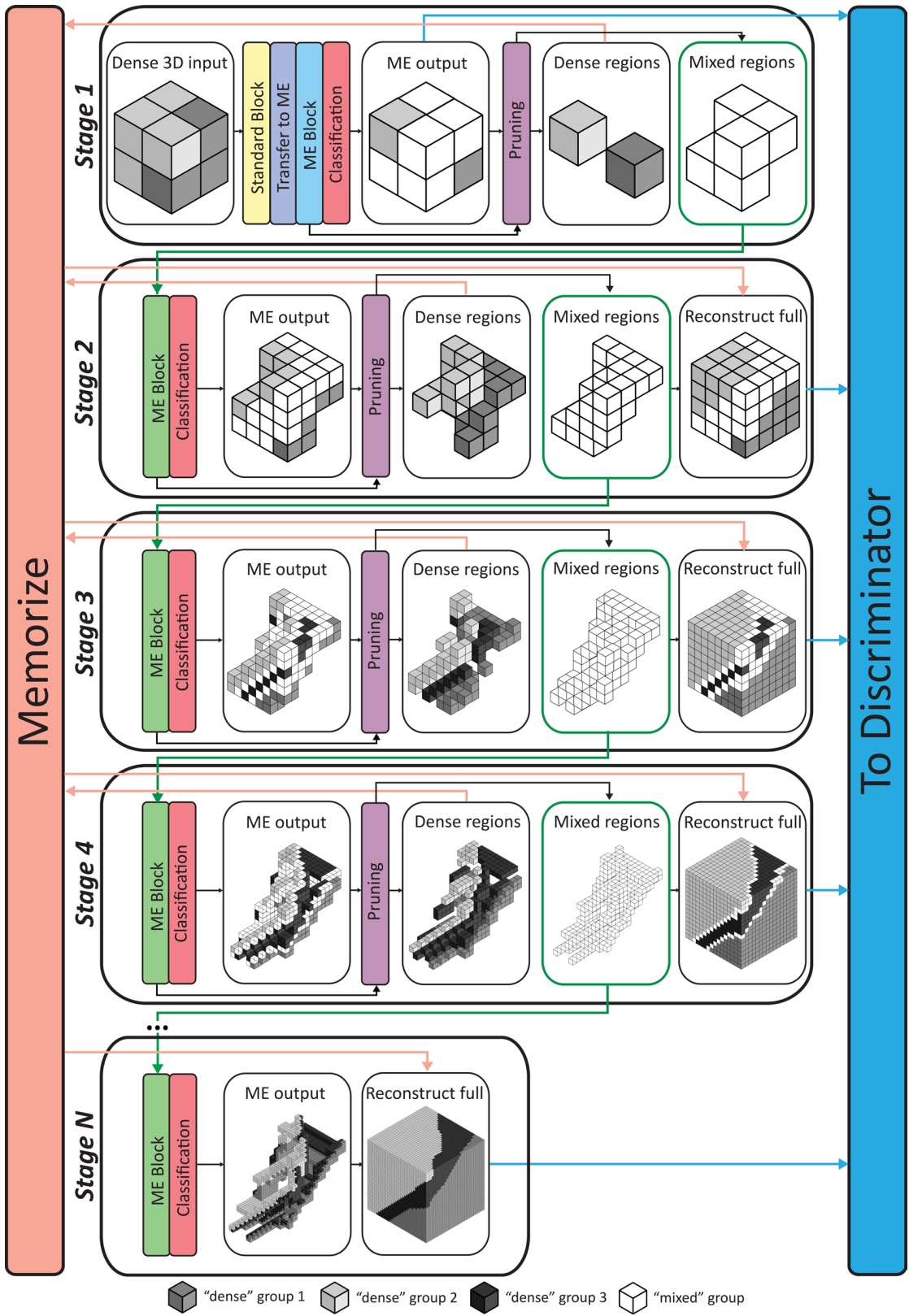

**Figure 4.** Schematic representation of the 3D Octree-Based Progressive Growing Generator (OB PG G) workflow. The figure illustrates the multi-stage refinement process, where input micro-CT images undergo

progressive subdivision and classification through N stages. At each stage, mixed regions are further refined, while dense regions are memorized and excluded from further computation. The final output is a Super-Resolution segmented 3D image with a 16x resolution enhancement. In our case, N = 5, but it can be a greater number, with the only limitation being the available GPU memory resources

The presented 3D Octree-Based Progressive Growing Generator (3D OB PG G) (**Table A1**) significantly reduces GPU memory consumption (**Figure 5**) compared to an analogous standard PyTorch implementation (**Table B1**). The model configurations and conservative memory consumption estimates for the presented 3D OB PG G, and its standard analog are provided in **Appendix A** and **Appendix B**, respectively.

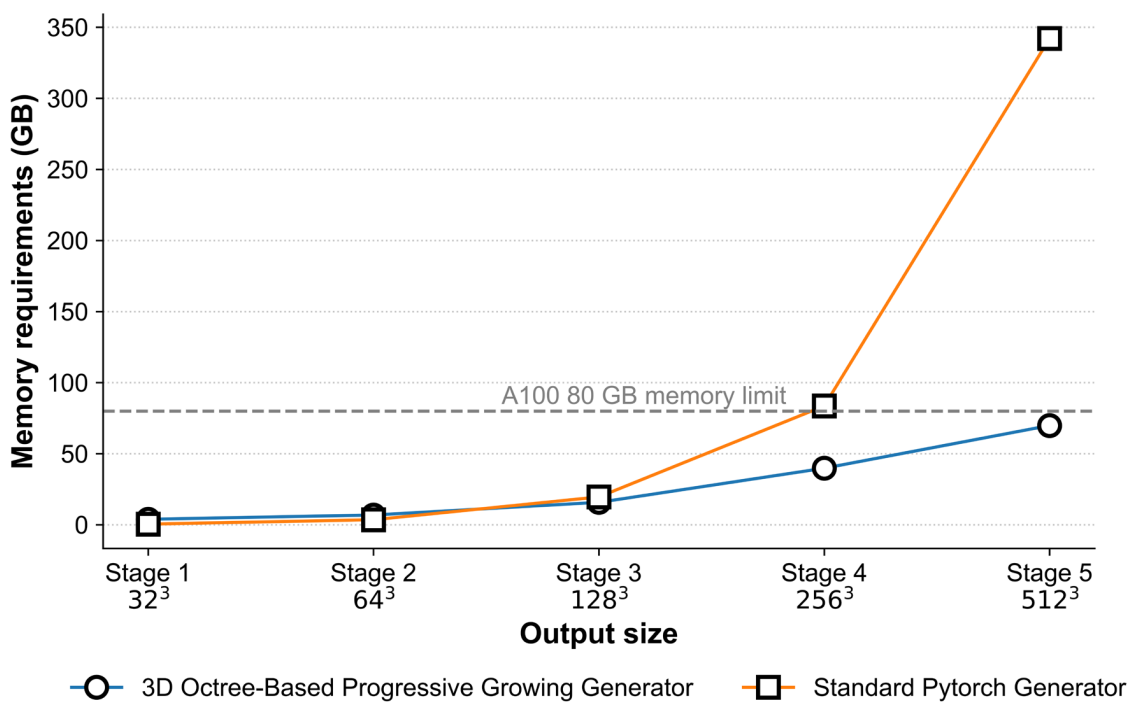

**Figure 5.** Comparison of memory consumption for the introduced memory-efficient 3D Octree-Based Progressive Growing Generator and its analog in the form of a dense Standard PyTorch Generator. With the presented configuration, the Standard PyTorch Generator produces Out Of Memory (OOM) error on Stage 4, making the Super-Resolution with a scale factor 16x impossible with the adequate model size. Alternatively, the presented memory-efficient 3D OB PG G optimization efficiently reduces the number of activations, significantly decreasing the memory consumption even for large output sizes

### 2.5. Progressive growing training dataset

To facilitate the training of each progressive growing stage (Stages 1-4), we generated additional training datasets derived from the original high-resolution segmented 2D LSM images. These datasets were constructed by systematically downscaling *dense* and *mixed* regions to match the resolution of each stage. The downscaling process involved applying a sliding window of progressively smaller patch sizes (16×16 pixels for Stage 1, 8×8 for Stage 2, 4×4 for Stage 3, and 2×2 for Stage 4) over the high-resolution images. Each patch was classified as *dense* (corresponding to a single phase such as pore, quartz, feldspar, or clay) if all pixels within the patch belonged to the same category. Patches containing two or more groups were labeled as *mixed*, ensuring that the dataset accurately reflected the hierarchical nature of the octree-based training process.

**Figure 6** illustrates this process. On the left, a high-resolution segmented image is divided into square patches of a predefined size corresponding to the resolution stage. Each patch is then evaluated for homogeneity. If all pixels within the patch belong to the same category (i.e., pore, quartz, feldspar, or clay) the patch is classified as *dense* and stored without further refinement. Conversely, if the patch contains multiple categories, it is labeled as *mixed* and marked for further subdivision in the subsequent resolution stages. The right side of **Figure 6** shows the corresponding classified patches, where *dense* regions remain unchanged, and *mixed* regions undergo further refinement.

**Figure 7** presents an example of a processed LSM image, demonstrating the progressive refinement of the training dataset across five stages. Each stage doubles the resolution while reducing the pixel size from 7 µm to 0.4375 µm. *Dense* regions (pore, quartz, feldspar, and clay) remain unchanged, while *mixed* regions are iteratively subdivided.

For Stage 5, the training dataset consisted of the original high-resolution segmented 2D images without downscaling, enabling the model to learn the final refinement at full resolution. To ensure compatibility with convolutional layers, we applied One-Hot-Encoding (OHE) after segmentation, enhancing the robustness of the training process.

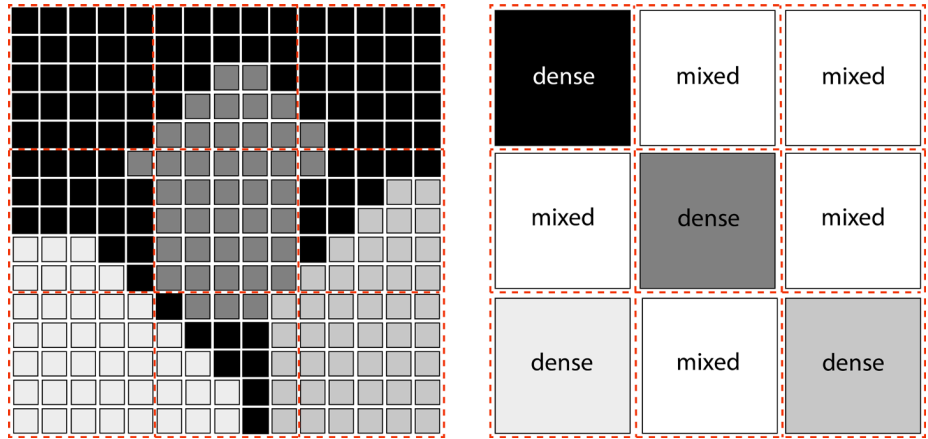

**Figure 6.** Illustration of the Progressive Growing dataset acquisition process. (Left) A high-resolution segmented 2D image is divided into square patches, with each patch evaluated for homogeneity. (Right) Patches are classified as *dense* if they contain a single mineral or pore phase, while those containing multiple phases are labeled as *mixed*. This classification determines whether a region requires further refinement in subsequent training stages. In this example, orange "patches" of size 5x5 pixels downscale the image of size 15x15 into an image of size 3x3.

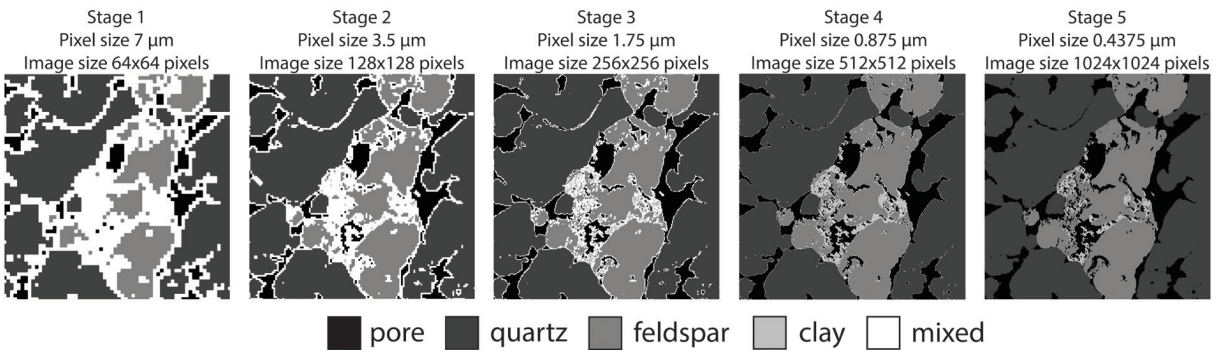

**Figure 7.** Example of the Progressive Growing High-Resolution training dataset across different stages. Each stage refines the resolution by a factor of two, progressively increasing the image size while decreasing the pixel size. Stage 1 begins with a pixel size of 7 µm (64×64 pixels), and by Stage 5, the resolution reaches 0.4375 µm (1024×1024 pixels). The classification of regions into pore (black), quartz (dark gray), feldspar (medium gray), clay (light gray), and mixed (white) is maintained throughout the progressive growing process. Each Stage is trained to differentiate "mixed" (white color) and "dense" (the rest of colors) nodes.

## 2.6. Progressive Growing Discriminator

The 2D Progressive Growing Discriminator (2D PG D) operates using standard PyTorch Conv2D layers, offering a significant reduction in memory consumption compared to Conv3D-based models. This efficiency allows it to function without requiring additional memory optimizations. To stabilize resolution transitions between stages, we adopted the progressive "fading-in" technique from Karras et al. (2017) , which is added to our 2D PG D network (**Figure 8**). In this approach, when a new stage begins, the newly introduced higher-resolution layer is initially treated as a residual block, with its contribution controlled by a weight parameter $\alpha$, which increases linearly from 0 to 1 within a transition period of 20 epochs. During this period, new layer processes actual High-Resolution (HR) and Super-Resolution (SR) images, while previously trained layers process the $\alpha$-weighted sum of down-sampled actual HR or SR images and output from new layer. This interpolation approach ensures a gradual resolution shift, preventing abrupt changes and stabilizing training across different scales. The architecture of the 2D PG D is presented in **Table C1**.

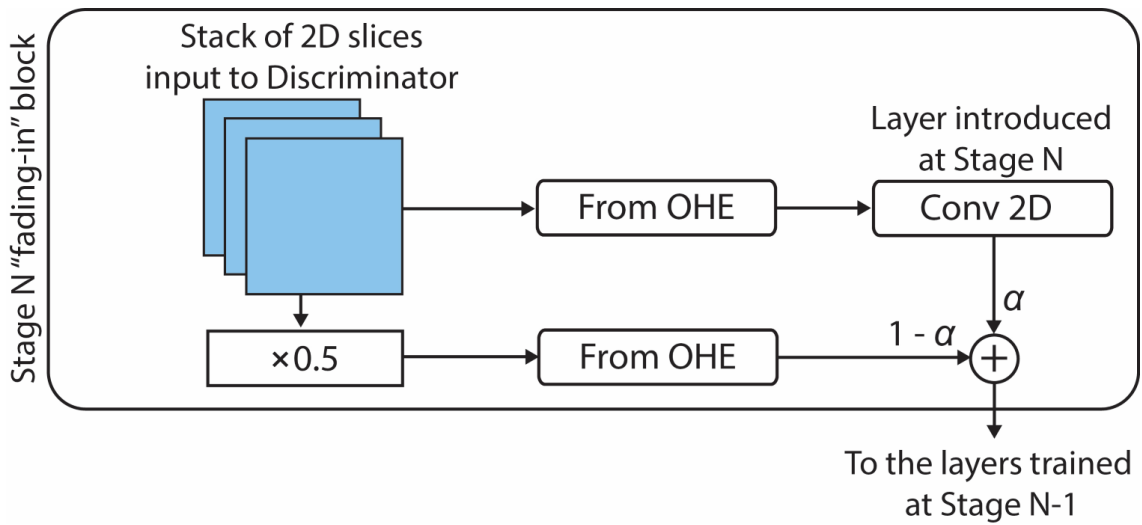

**Figure 8.** Illustration of the progressive "fading-in" mechanism used in the 2D PG D. At Stage N, a stack of 2D input images undergoes One-Hot-Decoding (From OHE) before being processed by a newly introduced Conv2D layer. Simultaneously, a down-sampled version (×0.5) of the input images is also One-Hot-Decoded. During the transition phase, the outputs from the new high-resolution layer and previously trained lower-resolution layers are interpolated using a weighting factor $\alpha$, which increases linearly from 0 to 1 for the first 20 Epochs.

### 2.7. Progressive Growing training algorithm

In the 3D OB PG G algorithm, the training process progressively enhances the resolution of *dense* segmented 3D super-resolution images through a hierarchical refinement approach. To achieve this, the training algorithm employs a Wasserstein Generative Adversarial Network with Gradient Penalty (WGAN-GP), extending upon the work of Ugolkov et al. (2025) and incorporating the methodology outlined by (Dahari et al., 2021). This strategy optimizes memory efficiency while preserving fine structural details in the reconstructed images.

The training process involves slicing the generated 3D super-resolution images along the x-, y-, and z-axes to create stacks of 2D images. These are then fed into the 2D Progressive Growing Discriminator (2D PG D), which compares them to randomly sampled segmented 2D images from the corresponding Progressive Growing training dataset (described in Section 2.5). The WGAN-GP loss for both the 3D OB PG G and 2D PG D models is computed based on the discriminator's outputs, following the approach outlined by (Gulrajani et al., 2017).

Additionally, a Mean Squared Error (MSE) loss is calculated between the Low-Resolution (LR) 3D pore space and the downsampled Super-Resolution (SR) 3D pore space to preserve the original structure. However, MSE loss is not applied to mineral groups, as one of the model's objectives is segmentation correction, and enforcing MSE loss on minerals would contradict this goal. The WGAN-GP and MSE losses are combined and used for backpropagation to update the parameters of both 3D OB PG G and 2D PG D.

Following training across all stages, the final 3D Super-Resolution images undergo post-processing using an iterative 3D median filter to eliminate minor artifacts. A schematic representation of the training process for a single stage N is shown in **Figure 9**, while the detailed pseudocode for the Progressive Growing training algorithm is provided in **Appendix E**.

**Figure 9** illustrates the Progressive Growing training algorithm for a single stage N, detailing the interaction between the 3D OB PG G and the 2D PG D. The process begins with the chunk of low-resolution 3D micro-CT image, which, along with noise input, is fed into the 3D OB PG G to generate an intermediate super-resolution 3D image at the current stage. To evaluate and refine this output, the SR 3D image is sliced along the x-, y-, and z-axes, producing 2D image stacks. These slices are then passed to the 2D PG D, which compares them to randomly selected high-resolution (HR) 2D images from the training dataset, determining their authenticity using a True/False Score loss. Simultaneously, a voxel-wise Mean Squared Error (MSE) loss is computed between the downsampled SR and LR pore spaces to ensure structural consistency.

The WGAN-GP loss and MSE loss are then backpropagated through the model, refining the generator and discriminator. This process is repeated for each stage, progressively increasing resolution while maintaining memory efficiency. Through this hierarchical approach, the model effectively enhances micro-CT images while addressing segmentation inaccuracies.

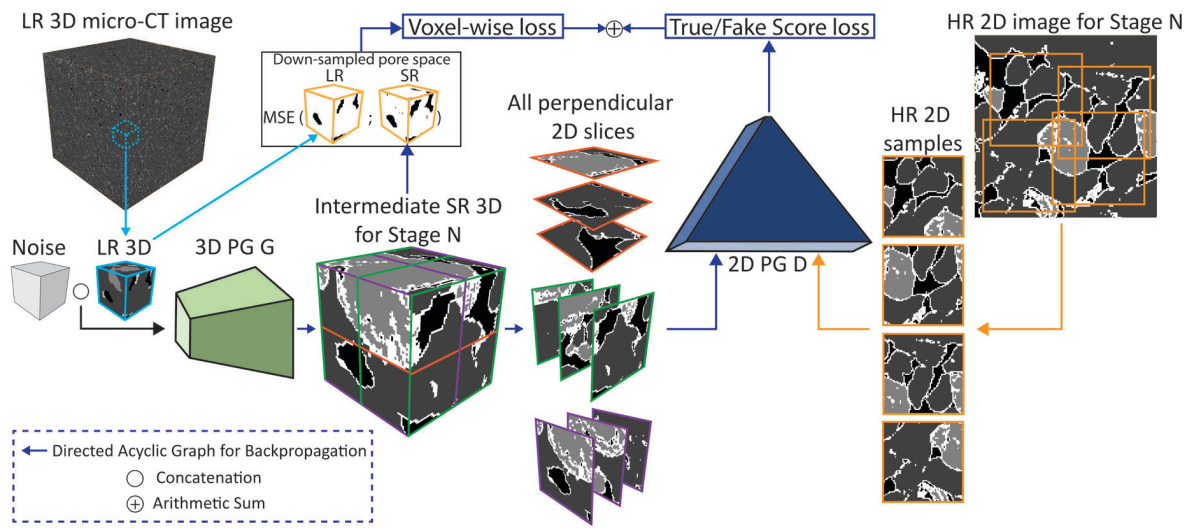

**Figure 9.** Schematic representation of the Progressive Growing training algorithm for a single Stage N. This particular example illustrates Stage 2. The 3D Progressive Growing Generator (3D PG G) generates an intermediate super-resolution (SR) 3D image from a low-resolution (LR) micro-CT image and noise input. The output is sliced into 2D images, which the 2D Progressive Growing Discriminator (2D PG D) evaluates against high-resolution (HR) training images, computing True/False Score and MSE losses. These losses are backpropagated, refining the model across stages while efficiently handling memory. Only mixed regions undergo super-resolution, while dense regions are reconstructed from stored data (see **Figure 4**).

### 2.8. Training details and hyperparameters

To further optimize memory usage, we employed training with Automatic Mixed Precision (AMP). AMP is a technique that optimizes deep neural network training by dynamically adjusting computations' precision using both 16-bit and 32-bit formats to improve performance and reduce memory usage. To overcome the occurrence of Not-a-Number (NaN), which is a common issue in Automatic Mixed Precision (AMP) training due to the reduced precision, we employed the AdamW optimizer for each stage. The optimizer was configured with an epsilon value of $10^{-4}$ to enhance numerical stability and a weight decay of 0.05. Additionally, the $\beta_1$ and $\beta_2$ parameters were set to 0.5 and 0.999, respectively, to optimize convergence and maintain training stability. For Stages 1-3, the learning rate was $10^{-4}$, and for Stages 4-5, the learning rate was $10^{-5}$. Also, we applied an epsilon value of $10^{-4}$ to the Instance Normalization (IN) and Batch Normalization (BN) layers. Each stage was trained for 100 epochs. To optimize memory usage, after training Stage N-1, the weights of the 3D PG G were frozen, and only the newly introduced layers were trained for Stage N. This approach significantly reduced the memory required for updating previously trained stages. Alternatively, we trained the 2D PG D entirely for all the Stages. The complete training process for all 5 stages took 71 hours, utilizing Distributed Data Parallel (DDP), batch size 2, on two 80 GB A100 GPUs.

## 3. Results

We evaluate the performance of the proposed algorithm using Berea sandstone as a case study. The model is not restricted to input volumes of $32^3$ voxels; instead, it can generate 3D images of arbitrary size by processing individual $32^3$ Low-Resolution (LR) sub-volumes and seamlessly stitching them together into a final Super-Resolution (SR) 3D image. This enables the preservation of larger-scale heterogeneities while maintaining a sufficiently large field of view (FOV).

As an example, we applied the 16× Super-Resolution algorithm to a 3D segmented LR micro-CT image of Berea sandstone with dimensions $256^3$ voxels (**Figure 10a**), producing a final segmented SR image with dimensions $4096^3$ voxels (**Figure 10b**). A qualitative comparison between the LR and SR images highlights significant improvements in the refinement of pore space and mineral boundaries (**Figure 10c-f**). The SR image reveals sub-micron pore structures and enhances intergranular connectivity, which were not visible in the LR dataset due to voxel size limitations. Additionally, the segmentation accuracy of feldspar, quartz, and clay is improved, leading to a more realistic representation of the rock microstructure.

The effectiveness of the proposed super-resolution algorithm was evaluated through quantitative comparisons of volume fraction, relative surface area, and two-point correlation functions for pore space across 2D HR, 3D SR, and 3D LR datasets. The volume fraction represents the ratio of voxels assigned to a specific group relative to the total voxel count in the region of interest. The relative surface area measures the proportion of interfacial faces between two groups relative to the total number of faces in the region. The two-point correlation function is a statistical metric that assesses the probability that two points, separated by a specific distance, belong to the same group. For each dataset, we randomly sampled 256 patches with dimensions $1024^2$, $1024^3$, and $64^3$ voxels, respectively, and calculated the metrics for each patch.

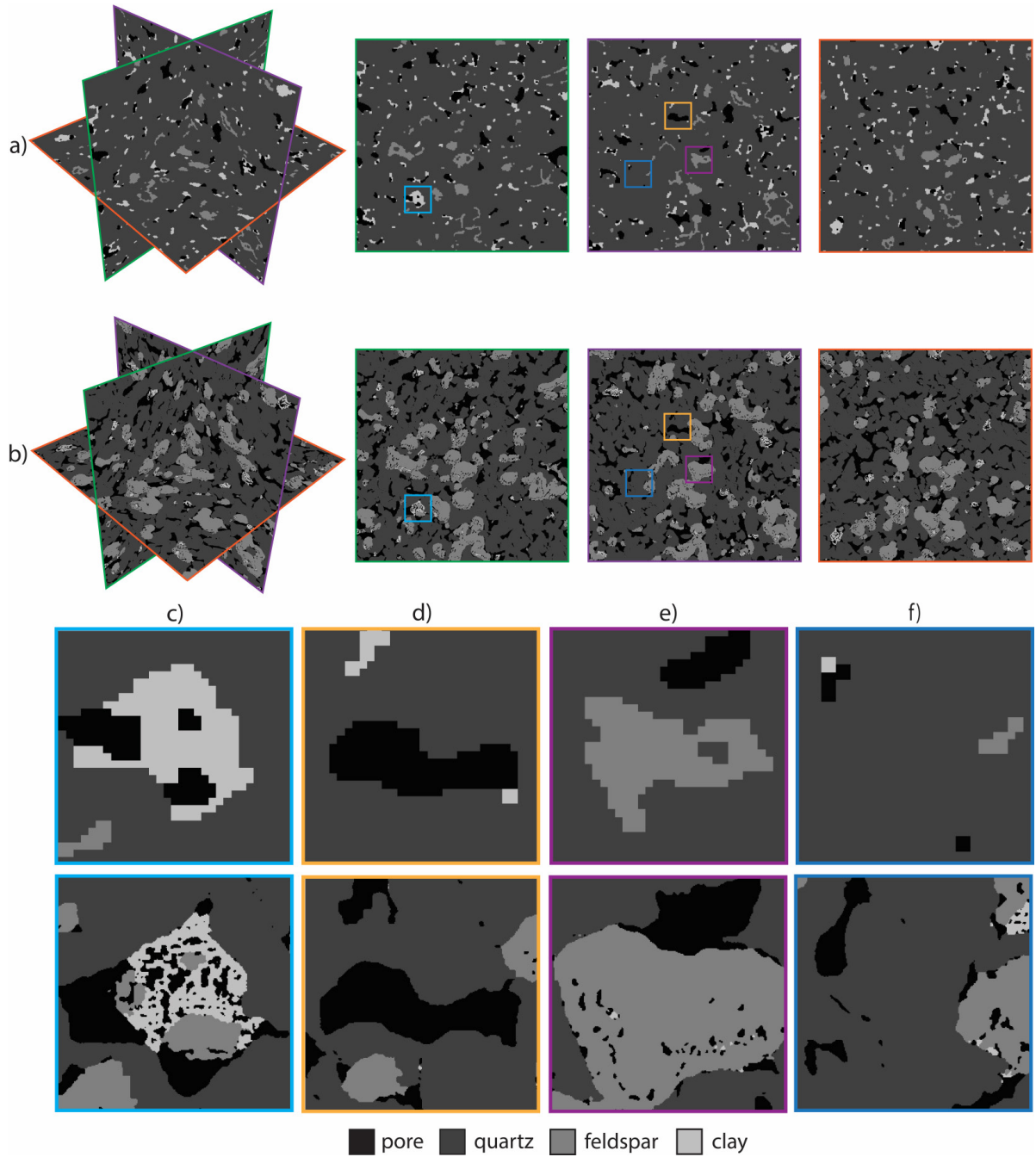

**Figure 10.** Comparison of LR and SR segmented 3D micro-CT images of Berea sandstone. (a) and (b) show orthogonal slices of the 256³ voxel LR image and its corresponding 4096³ voxel SR (16x) image generated using the proposed algorithm. (c-f) provide zoomed-in regions from the LR and SR images, illustrating the refinement of pore space and mineral boundaries. The SR image reveals finer structural details, improved segmentation accuracy, and enhanced pore connectivity compared to the LR images.

**Figure 11a** presents the volume fraction distribution for each classified phase (pore, quartz, feldspar, and clay). The 3D SR dataset closely aligns with the 2D HR dataset, demonstrating a substantial improvement over the 3D LR dataset, particularly in the pore space fraction, which was significantly underestimated in LR due to resolution limitations. Additionally, **Figure 11b** compares the relative surface area between different phase boundaries. To further assess spatial continuity, **Figure 12** illustrates the two-point correlation function for pore space, comparing probability distributions at increasing lag distances. The 3D SR dataset exhibits a correlation trend that closely follows the HR dataset, reinforcing the model's ability to recover fine-scale heterogeneities and accurately reconstruct complex pore structures.

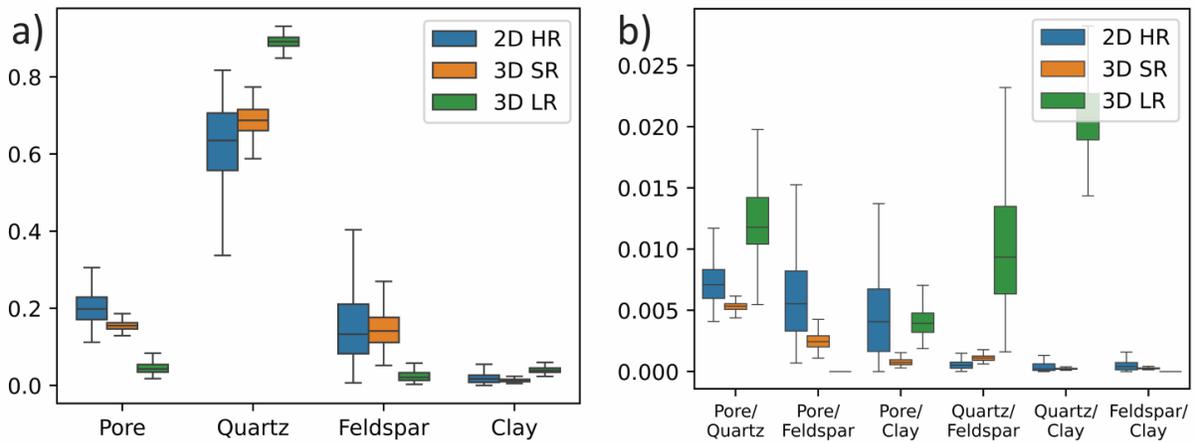

**Figure 11.** Box plots for (a) Volume fraction comparison for pore, quartz, feldspar, and clay phases in 2D HR, 3D SR, and 3D LR datasets. The 3D SR dataset closely aligns with the 2D HR reference. (b) Relative surface area for different phase boundaries, showing that 3D SR significantly reduces segmentation artifacts and better approximates the HR dataset compared to 3D LR.

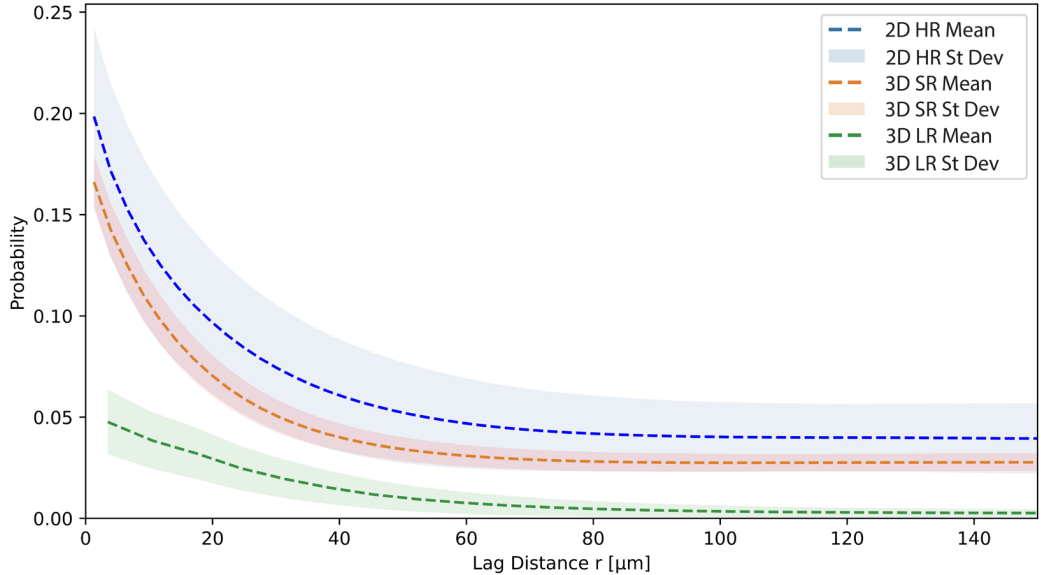

**Figure 12.** Two-point correlation function analysis of pore space across different datasets. The plot compares the probability of two points, separated by a given lag distance, belonging to the same phase in 2D HR, 3D SR, and 3D LR images. The 3D SR dataset exhibits a closer match to the 2D HR reference. The shaded regions represent standard deviations across sampled patches.

To further illustrate the impact of the Super-Resolution approach on the pore space structure, we extracted cubic sub-volumes of $64^3$ and $1024^3$ voxels from the central region of both the 3D LR and 3D SR volumes shown in **Figure 10**. These extracted volumes were visualized using Voxellized Rendering in PerGeos using the Voxellized Rendering functionality of PerGeos (Thermo Fisher, 2021) to highlight the structural improvements introduced by the super-resolution process. **Figure 13** presents a comparative 3D visualization of the pore space, where panel (a) shows the Low-Resolution image, characterized by coarse, blocky features, while panel (b) depicts the Super-Resolution image, revealing significantly more intricate and continuous pore structures. The improved resolution allows for a more detailed representation of fine-scale porosity.

We performed Pore Network Extraction (PNE) on both datasets and computed the Pore Size Distribution (PSD), as shown in **Figure 14**. The histogram compares the equivalent radius of pores in the 3D Super-Resolution and 3D Low-Resolution volumes. The Low-Resolution dataset exhibits a predominance of larger pores, with a notable peak around 70 μm, indicating an underrepresentation of finer-scale porosity. In contrast, the Super-Resolution dataset resolves a broader range of pore sizes, particularly enhancing the representation of small-scale pores below 25 μm. This expanded distribution aligns more closely with high-resolution imaging results and provides a more representative rock's pore structure.

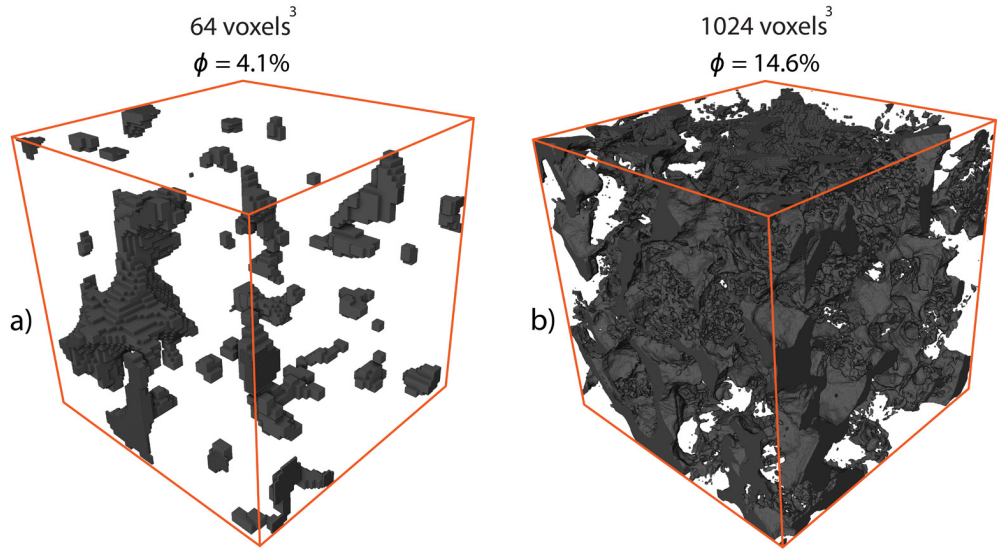

**Figure 13.** 3D visualization of the extracted pore space sub-volumes from the (a) Low-Resolution consisting of $64^3$ voxels and (b) Super-Resolution consisting of and $1024^3$ voxels.

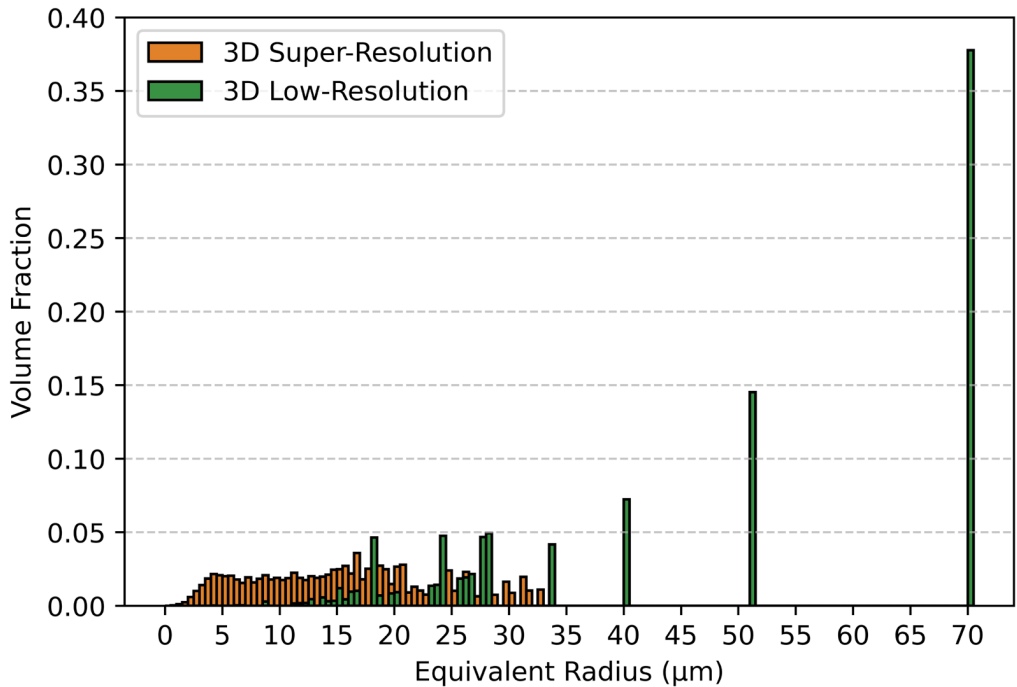

**Figure 14.** Comparison of equivalent radius between the Low-Resolution and Super-Resolution datasets. The Super-Resolution dataset (orange) exhibits a broader distribution of smaller pores, indicating enhanced resolution and the ability to capture fine-scale porosity. In contrast, the Low-Resolution dataset (green) is dominated by larger pore structures, failing to resolve sub-micron features.

Applying the Octree-Based Progressive Growing approach was critical in achieving computational efficiency while enabling 3D Super-Resolution. By selectively processing only "mixed" nodes and storing "dense" regions (pore, quartz, feldspar, and clay), the model significantly reduced memory consumption and computational costs (**Figure 5**), making large-scale high-resolution reconstruction feasible. As shown in **Figure 15**, the algorithm progressively refines the input $32^3$ voxel sub-volume, where Stage 1 starts with a large proportion of mixed nodes due to the coarse resolution. With each successive stage, Stages 2-4 incrementally refine the segmentation, subdividing only complex regions while preserving homogeneous areas. By Stage 5, the model reconstructs a fully segmented $512^3$ voxel resolution image, followed by a median filter smoothing operation to further enhance segmentation accuracy and reduce artifacts. The quantitative analysis in **Table 2** demonstrates the efficiency of this approach, showing a substantial decrease in the proportion of mixed nodes as the resolution increases. This selective refinement process ensures that computational resources are dedicated to the most critical regions, optimizing memory usage while maintaining high fidelity in the final 3D Super-Resolution output.

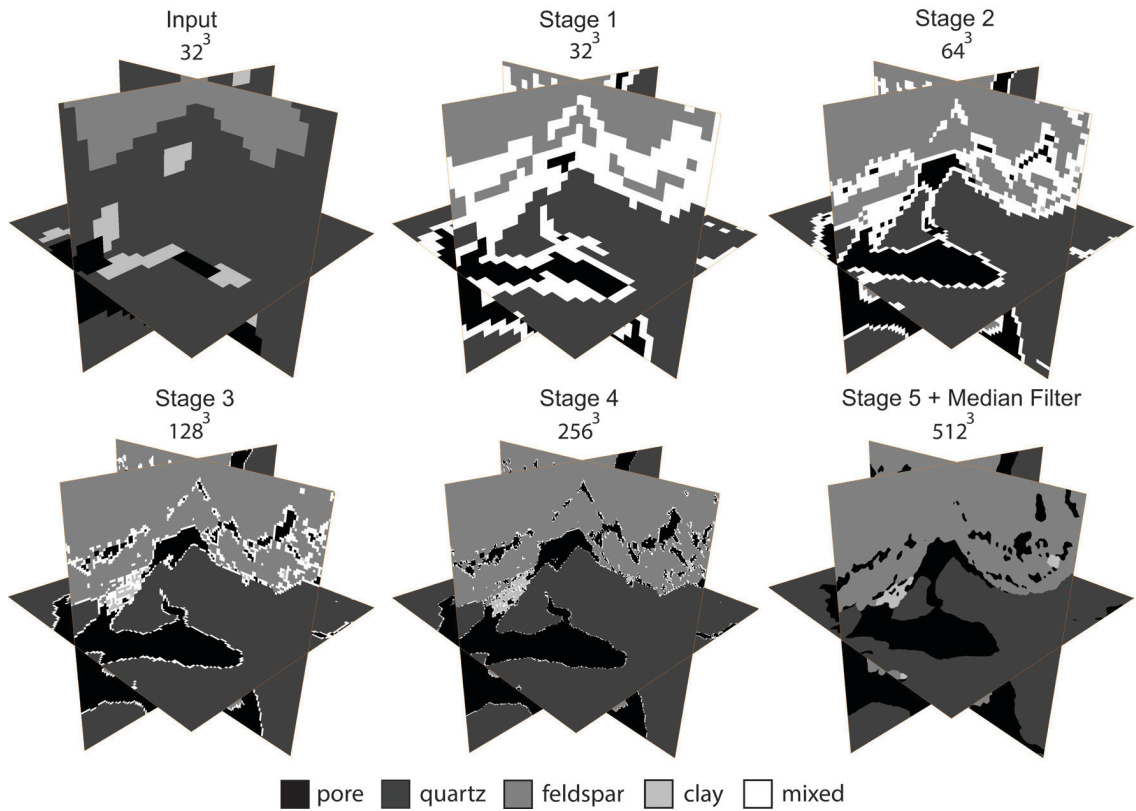

**Figure 15.** Visualization of the Octree-Based Progressive Growing refinement process across the 5 stages. The input $32^3$ voxel sub-volume undergoes progressive subdivision, where each stage selectively refines "mixed" nodes while storing "dense" regions. As the resolution increases from Stage 1 to Stage 5, the segmentation becomes more detailed.

**Table 2.** Proportion of standard voxels and Minkowski Engine (ME) *dense* and *mixed* nodes on each Stage output, demonstrating the efficiency of the octree structure in reducing computational load.

| Stage | Output size in standard format | Total number of voxels in standard format | Number of *dense* nodes in ME format | Number of *mixed* nodes in ME format | Percentage of *mixed* nodes in total number of standard voxels |
|---|---|---|---|---|---|
| 1 | $32^3$ | 32,768 | 23,238 | 9,530 | 29.1 |
| 2 | $64^3$ | 262,144 | 36,103 | 40,137 | 15.3 |
| 3 | $128^3$ | 2,097,152 | 171,071 | 150,025 | 7.2 |
| 4 | $256^3$ | 16,777,216 | 743,712 | 456,488 | 2.7 |
| 5 | $512^3$ | 134,217,728 | 3,651,904 | 0 | 0 |

## 4. Discussion

This study demonstrates a 16× resolution enhancement of segmented 3D micro-CT images, improving the voxel size resolution from 7 µm to 0.44 µm. This resolution makes it possible to identify the presence of features with an effective radius of size 0.87 µm while quantifying in detail the features with an effective radius of size 2.19 µm. The proposed algorithm addresses three key challenges in micro-CT image super-resolution, consisting of:

*1. Incorporation of sub-micron features:* The model resolves fine-scale structures, particularly in pore space and mineral boundaries, which were previously undetected due to the limitations of low-resolution imaging. In **Figure 10c-e**, the algorithm refines homogeneous regions of clay and feldspar, introducing realistically structured mineral grains and intragranular porosity. Similarly, in **Figure 10f**, quartz grains are better delineated, with the emergence of intergranular porosity. This structural refinement is further quantified in **Figure 11a**, where the pore volume fraction increases from 5% to 17%, while quartz volume decreases due to the introduction of intergranular pores. Additionally, improvements in pore connectivity are evident in **Figure 12**, where the two-point correlation function of the super-resolved dataset aligns more closely with the high-resolution benchmark. The incorporation of additional flow pathways (**Figure 13**) further supports the algorithm's effectiveness in improving the hydraulic characterization of digital rock models. The PNE analysis reinforces these findings, demonstrating a significant increase in the number of detected pores and throats (**Figure 14**). In the low-resolution dataset, only 50 pores and 9 throats were identified, whereas the super-resolved dataset revealed 14,919 pores and 18,475 throats, including a substantial proportion of sub-micron features that were previously unresolved. This enhancement is particularly critical for fluid flow simulations, where accurate pore-scale characterization directly impacts permeability and transport predictions. Furthermore, the improved surface resolution of segmented phases enhances surface roughness calculations and simulations involving grain-fluid interactions, such as Nuclear Magnetic Resonance (NMR) Random Walk modeling (Li et al., 2023). This is quantitatively verified in **Figure 11b**, where the relative surface area measurements for the super-resolved dataset are significantly closer to the high-resolution ground truth.

*2. Correction of segmentation inaccuracies*: Another key advantage of the proposed algorithm is its ability to correct segmentation errors in minerals with similar electron densities, specifically feldspars and quartz in our case. In **Figure 10a,b,e**, regions containing feldspar traces are corrected to match realistic mineral structures corresponding to those observed in 2D High-Resolution microscope images (**Figure 2**). This could be observed in the increase of feldspar volume fraction in **Figure 11a**, where the 17% mean value for Super-Resolution aligns well with the XRD data (Ugolkov et al., 2025). Improved mineral differentiation has direct implications for digital rock physics simulations, particularly those modeling elastic properties and electrical conductivity.

*3. Computational Efficiency and Memory Optimization:* The proposed Octree-Based Progressive Growing Super-Resolution algorithm achieves superior resolution scaling (16x) while maintaining computational efficiency. This improvement is made possible by three key factors. First, the adoption of a memory-efficient Progressive Growing Octree structure, which processes only a small fraction of the "mixed" nodes between Stages (**Figure 15** and **Table 2**). For instance, in the presented case, Stage 4 transferred only 2.7% of the nodes to Stage 5, allowing the model to accommodate a larger number of parameters instead. Second, the Progressive Growing training approach enables Stage-wise training while freezing the parameters of earlier Stages, thereby freeing computational resources for the later Stages. Finally, the proposed modification integrates effectively with Automatic Mixed Precision (AMP). Implementing AMP training further minimizes memory usage, allowing larger model sizes to fit within a single GPU without performance degradation.

## 3. Conclusion

We introduced a memory-efficient Super-Resolution algorithm for segmented 3D micro-CT images, achieving a 16× resolution enhancement from 7 µm to 0.44 µm and successfully tested it on Berea sandstone. This approach resolves sub-micron features as well as corrects segmentation inaccuracies. The Octree-Based Progressive Growing implementation leverages the dense voxelized structure of micro-CT images, effectively distinguishing between coarse-detailed and fine-detailed regions. By selectively processing only fine-detailed regions in later stages, the algorithm drastically reduces memory consumption and training time compared to standard dense PyTorch 3D convolutional implementations. A key advantage of this method is its independent and scalable training process, which utilizes unpaired 3D micro-CT and 2D Laser Scanning Microscope (LSM) images. This flexibility simplifies dataset acquisition while ensuring high-fidelity reconstructions. The training was conducted using a 3D Octree-Based Progressive Growing Deep Convolution Wasserstein Generative Adversarial Network with Gradient Penalty (3D OB PG DC WGAN-GP), incorporating 3D Generator and 2D Discriminator models. After training, the model can generate high-resolution 3D Super-Resolved images of arbitrary sizes by processing and stitching together $32^3$ sub-volumes, maintaining the original field of view of a micro-CT scan while achieving the resolution of an LSM.

This advancement has the potential to improve Digital Rock Physics (DRP) simulations, enabling more accurate fluid flow. Moving forward, we will focus on two key areas: (1) evaluating the algorithm across different rock types, including tight sandstones, where resolution limitations pose significant challenges, and (2) assessing the impact of Super-Resolution on various DRP simulations to enhance their predictive accuracy. By significantly improving both efficiency and resolution, this method paves the way for more detailed and computationally feasible digital rock analyses, making high-fidelity 3D Super-Resolution imaging a practical tool in geosciences and material research.

## 4. Code availability

The *"x16-Octree-Based-Super-Resolution"* code is available for use and can be accessed through the GitHub repository at https://github.com/EvgenyUgolkov/x16-Octree-Bassed-Super-Resolution. For inquiries, users may contact Evgeny Ugolkov at evgeny.ugolkov@kaust.edu.sa. The implementation requires two NVIDIA A100 80 GB GPUs and 450 GB of RAM for optimal performance. The code is written in Python and requires segmentation software for dataset preparation, such as Avizo. The total program size, including the training dataset and evaluation volume, is 376 MB.

## 5. Acknowledgements

We express our deep gratitude to Dr. Mohsin Ahmed Shaikh and Dr. Rooh Khurram from the KAUST Supercomputing Core Laboratory (KSL) for their help with the work on the IBEX supercomputer cluster and Minkowski Engine library installation. We extend our thanks to Mr. Domingo A. Lattanzi Sanchez for his invaluable guidance on optimizing micro-CT parameters for high-quality image acquisition. We also thank Dr. Jeffery Carpenter for his support in preparing the thin sections. Additionally, we are grateful to Dr. Ivan Skorohodov for the insightful discussions on the StyleGAN2ADA algorithm and its application to segmented LSM images.

## 6. Declaration of Generative AI and AI-assisted technologies in the writing process

During the preparation of this work, the authors used ChatGPT 4o to paraphrase some parts of this work in a formal scientific style to improve readability and language. After using this tool, the authors reviewed and edited the content as needed and take full responsibility for the content of the publication.

## 7. Declaration of competing interest

The authors declare that they have no known competing financial interests or personal relationships that could have appeared to influence the work reported in this paper.

**Appendix A:** 3D Octree-Based Progressive Growing Generator Architecture and Memory Consumption

**Table A1.** Architecture of the 3D Octree-Based Progressive Growing Generator (3D OB PG G), including the layer configurations, activation functions, sizes of output, number of trainable parameters, and Stage boundaries. The colors correspond to **Figure 4**.

| Layer | Activation | Max Output Shape | Params | Stage |
|---|---|---|---|---|
| Dense Input | - | 5 x 32 x 32 x 32 | - | |
| Conv3d | - | 512 x 32 x 32 x 32 | 320.5k | |
| InstanceNorm3d | ReLu | 512 x 32 x 32 x 32 | 1k | |
| Conv3d | - | 512 x 32 x 32 x 32 | 7.1M | |
| InstanceNorm3d | ReLu | 512 x 32 x 32 x 32 | 1k | |
| Transfer to ME format | - | 512 x 32 x 32 x 32 | - | 1 |
| ME Conv3d | - | 512 x 32 x 32 x 32 | 7.1M | |
| ME BatchNorm3d | ME ReLu | 512 x 32 x 32 x 32 | 1k | |
| ME Conv3d | - | 5 x 32 x 32 x 32 | 2.5k | |
| ME Softmax | - | 5 x 32 x 32 x 32 | - | |
| ME Pruning | - | 512 x 32 x 32 x 32 | - | |
| ME ConvTranspose3d | - | 512 x 64 x 64 x 64 | 2.1M | |
| ME BatchNorm3d | ME ReLu | 512 x 64 x 64 x 64 | 1k | |
| ME Conv3d | - | 512 x 64 x 64 x 64 | 7.1M | |
| ME BatchNorm3d | ME ReLu | 512 x 64 x 64 x 64 | 1k | |
| ME Conv3d | - | 512 x 64 x 64 x 64 | 7.1M | 2 |
| ME BatchNorm3d | ME ReLu | 512 x 64 x 64 x 64 | 1k | |
| ME Conv3d | - | 5 x 64 x 64 x 64 | 2.5k | |
| ME Softmax | - | 5 x 64 x 64 x 64 | - | |
| ME Pruning | - | 512 x 64 x 64 x 64 | - | |
| ME ConvTranspose3d | - | 512 x 128 x 128 x 128 | 2.1M | |
| ME BatchNorm3d | ME ReLu | 512 x 128 x 128 x 128 | 1k | |
| ME Conv3d | - | 256 x 128 x 128 x 128 | 3.6M | |
| ME BatchNorm3d | ME ReLu | 256 x 128 x 128 x 128 | 512 | |
| ME Conv3d | - | 256 x 128 x 128 x 128 | 1.77M | 3 |
| ME BatchNorm3d | ME ReLu | 256 x 128 x 128 x 128 | 512 | |
| ME Conv3d | - | 5 x 128 x 128 x 128 | 1.3k | |
| ME Softmax | - | 5 x 128 x 128 x 128 | - | |
| ME Pruning | - | 256 x 128 x 128 x 128 | - | |
| ME ConvTranspose3d | - | 256 x 256 x 256 x 256 | 524.5k | |
| ME BatchNorm3d | ME ReLu | 256 x 256 x 256 x 256 | 512 | |
| ME Conv3d | - | 128 x 256 x 256 x 256 | 884.6k | |
| ME BatchNorm3d | ME ReLu | 128 x 256 x 256 x 256 | 256 | |
| ME Conv3d | - | 128 x 256 x 256 x 256 | 442.5k | 4 |
| ME BatchNorm3d | ME ReLu | 128 x 256 x 256 x 256 | 256 | |
| ME Conv3d | - | 5 x 256 x 256 x 256 | 645 | |
| ME Softmax | - | 5 x 256 x 256 x 256 | - | |
| ME Pruning | - | 128 x 256 x 256 x 256 | - | |
| ME ConvTranspose3d | - | 128 x 512 x 512 x 512 | 131k | |
| ME BatchNorm3d | ME ReLu | 128 x 512 x 512 x 512 | 256 | |
| ME Conv3d | - | 64 x 512 x 512 x 512 | 221.9k | |
| ME BatchNorm3d | ME ReLu | 64 x 512 x 512 x 512 | 128 | |
| ME Conv3d | - | 64 x 512 x 512 x 512 | 111k | 5 |
| ME BatchNorm3d | ME ReLu | 64 x 512 x 512 x 512 | 128 | |
| ME Conv3d | - | 5 x 512 x 512 x 512 | 325 | |
| ME Softmax | - | 5 x 512 x 512 x 512 | - | |

Total trainable parameters **40.5M**

The GPU memory consumption for the presented 3D OB PG G was traced with the Weights and Biases toolset (Biewald, 2020), as demonstrated in **Figure A1**. Due to the Octree-Based implementation, different epochs process a different number of "mixed" and "dense" nodes, resulting in slight variations in memory consumption. For **Figure 5**, we plotted the maximum values for each Stage.

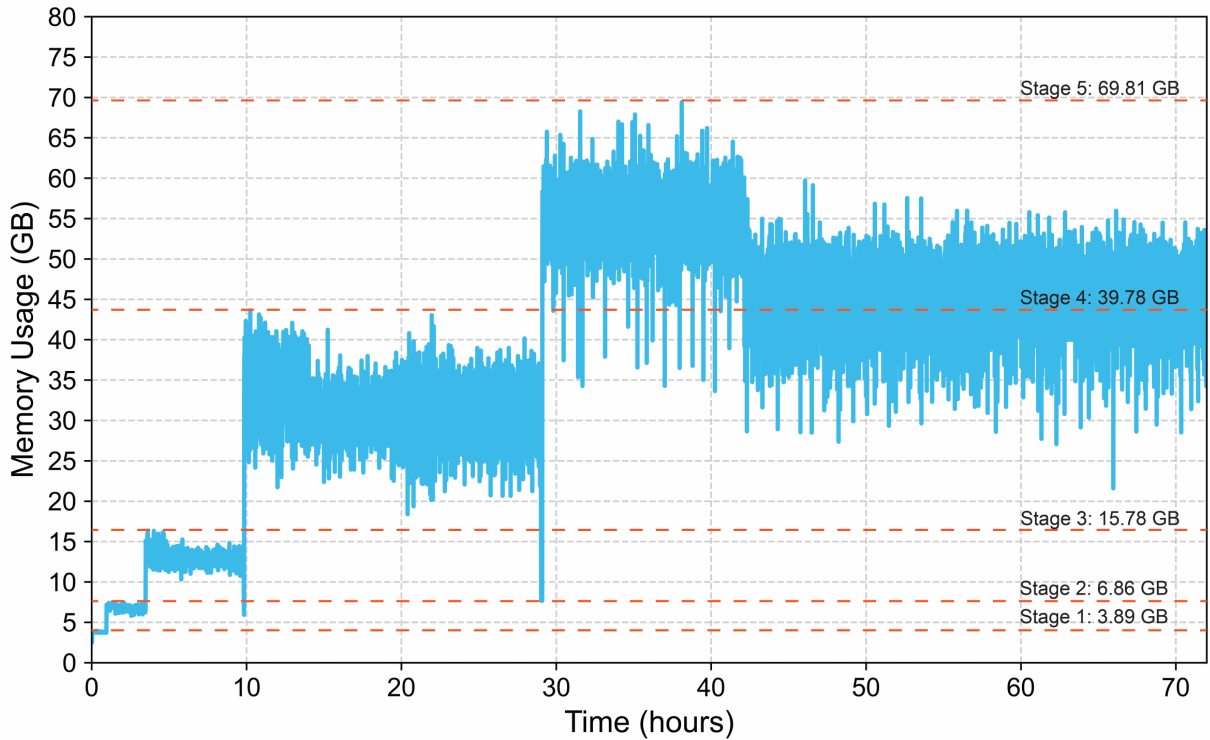

**Figure A1.** Memory consumption for the 3D Octree-Based Progressive Growing Generator for all Stages for batch size 1

**Appendix B:** Architecture and conservative Memory Consumption Estimation for Standard PyTorch Generator analogous to the 3D OB PG G introduced in this work

**Table B1**. Architecture for Standard PyTorch Generator analogous to the 3D OB PG G, including the layer configurations, activation functions, sizes of output, number of trainable parameters, incremental and cumulative memory consumption, and Stage outline.

| Layer name | Activation | Output size | Parameters | Memory per layer, GB | Cumulative memory, GB | Stage |
|---|---|---|---|---|---|---|
| Conv3D | - | 512×32×32×32 | 320512 | 0.06 | 0.064 | 1 |
| InstNorm3D | ReLU | 512×32×32×32 | 1024 | 0.06 | 0.127 | 1 |
| Conv3D | - | 512×32×32×32 | 7078400 | 0.10 | 0.229 | 1 |
| InstNorm3D | ReLU | 512×32×32×32 | 1024 | 0.06 | 0.291 | 1 |
| Conv3D | - | 512×32×32×32 | 7078400 | 0.10 | 0.393 | 1 |
| BatchNorm3D | ReLU | 512×32×32×32 | 1024 | 0.06 | 0.456 | 1 |
| ConvTr3D | - | 512×64×64×64 | 2097664 | 0.51 | 0.968 | 2 |
| BatchNorm3D | ReLU | 512×64×64×64 | 1024 | 0.50 | 1.468 | 2 |
| Conv3D | - | 512×64×64×64 | 7078400 | 0.54 | 2.007 | 2 |
| BatchNorm3D | ReLU | 512×64×64×64 | 1024 | 0.50 | 2.507 | 2 |
| Conv3D | - | 512×64×64×64 | 7078400 | 0.54 | 3.047 | 2 |
| BatchNorm3D | ReLU | 512×64×64×64 | 1024 | 0.50 | 3.547 | 2 |
| ConvTr3D | - | 512×128×128×128 | 2097664 | 4.01 | 7.558 | 3 |
| BatchNorm3D | ReLU | 512×128×128×128 | 1024 | 4.00 | 11.558 | 3 |
| Conv3D | - | 256×128×128×128 | 3539200 | 2.02 | 13.578 | 3 |
| BatchNorm3D | ReLU | 256×128×128×128 | 512 | 2.00 | 15.578 | 3 |
| Conv3D | - | 256×128×128×128 | 1769728 | 2.01 | 17.588 | 3 |
| BatchNorm3D | ReLU | 256×128×128×128 | 512 | 2.00 | 19.588 | 3 |
| ConvTr3D | - | 256×256×256×256 | 524544 | 16.00 | 35.591 | 4 |
| BatchNorm3D | ReLU | 256×256×256×256 | 512 | 16.00 | 51.591 | 4 |
| Conv3D | - | 128×256×256×256 | 884864 | 8.00 | 59.596 | 4 |
| BatchNorm3D | ReLU | 128×256×256×256 | 256 | 8.00 | 67.596 | 4 |
| Conv3D | - | 128×256×256×256 | 442496 | 8.00 | 75.599 | 4 |
| BatchNorm3D | ReLU | 128×256×256×256 | 256 | 8.00 | 83.599 | 4 |
| ConvTr3D | - | 128×512×512×512 | 131200 | 64.00 | 147.599 | 5 |
| BatchNorm3D | ReLU | 128×512×512×512 | 256 | 64.00 | 211.599 | 5 |
| Conv3D | - | 64×512×512×512 | 221248 | 32.00 | 243.600 | 5 |
| BatchNorm3D | ReLU | 64×512×512×512 | 128 | 32.00 | 275.600 | 5 |
| Conv3D | - | 64×512×512×512 | 110656 | 32.00 | 307.601 | 5 |
| BatchNorm3D | ReLU | 64×512×512×512 | 128 | 32.00 | 339.601 | 5 |
| Conv3D | Softmax | 5×512×512×512 | 325 | 2.50 | 342.101 | 5 |

For the conservative calculations of memory consumption, we were guided by the consideration that for the training, the total GPU memory is consumed by the Activations and Model Parameters.

Activations:

1. Each layer output activation must be stored in memory during the forward pass
2. During backpropagation, gradients are computed for each of these activations, doubling their memory footprint

Model Parameters:

1. All layer parameters, namely weights and biases, are stored
2. Additionally, the Adam optimizer stores two auxiliary tensors per parameter (1st and 2nd moments), resulting in 3× parameter memory

This way, the total memory usage can be calculated as follows:

$$M_{Total} = M_{Activations} + M_{Params} = (2 \times \sum_{i=1}^{L} Elements(A_i) + 3 \times \sum_{i=1}^{L} Elements(W_i)) \times \text{BytesPerElement},$$

where

$L$ is the total number of layers,

$A_i$ is the number of Activations per layer $i$, which is a multiplication of output size dimensions,

$W_i$ is the number of parameters per layer $i$,

BytesPerElement is 2 for Mixed Precision implementation (float16 format);

Upon such analysis, we calculated the incremental and cumulative memory consumption for the analogous standard PyTorch Generator (**Table B1**) and compared the results with the requirements for the introduced 3D Octree-Based Memory-Efficient Progressive Growing Generator in **Figure 5**.

**Appendix C:** 2D Progressive Growing Discriminator Architecture

**Table C1.** Architecture of the 2D Progressive Growing Discriminator (2D PG D), specifying the layer configurations, kernel sizes, activation functions, output shape, and the number of training parameters across different training Stages.

| Layer | Kernel size | Activation | Output Shape | Params | Stage |
|---|---|---|---|---|---|
| Input | - | - | 5 x 512 x 512 | - | |
| Conv2d | 3 x 3 | ReLU | 64 x 256 x 256 | 2.9k | 5 |
| Conv2d | 3 x 3 | ReLU | 128 x 128 x 128 | 74k | 4 |
| Conv2d | 3 x 3 | ReLU | 256 x 64 x 64 | 295k | 3 |
| Conv2d | 3 x 3 | ReLU | 512 x 32 x 32 | 1.2M | 2 |
| Conv2d | 3 x 3 | ReLU | 512 x 16 x 16 | 2.36M | |
| Conv2d | 3 x 3 | ReLU | 512 x 8 x 8 | 2.36M | |
| Conv2d | 3 x 3 | ReLU | 512 x 4 x 4 | 2.36M | 1 |
| Conv2d | 4 x 4 | ReLU | 512 x 1 x 1 | 4.2M | |
| Fully connected | - | Linear | 1 x 1 x 1 | 513 | |
| | | | Total trainable parameters | **12.8M** | |

**Appendix D**: 16x 3D Octree-Based Progressive Growing Generator (OB PG G) algorithm

**Input:**

- $X$ – the number of Stages;
- $O$ – the Stage in training;
- $LR$ – the 3D Low-Resolution training dataset;
- `standard conv block` – a sequence of PyTorch Conv3d, InstanceNorm3d, and ReLU layers;
- `Minkowski Engine format` – a function which takes a standard PyTorch tensor and transfers it to the Minkowski Engine (ME) tensor;
- `Minkowski Engine conv` – a sequence of MinkowskiConvolution, MinkowskiBatchNorm, and MinkowskiReLU layers;
- `Minkowski Engine conv trans` – a sequence of MinkowskiGenerativeConvolutionTranspose, MinkowskiConvolution, MinkowskiBatchNorm, and MinkowskiReLU layers;
- `Classification` – is the MinkowskiConvolution with kernel size = 1 and output number of channels equal to number of phases plus one (all the minerals, pore space, and the mixed group);
- `Softmax` – is the MinkowskiSoftmax(dim=1) for predicting the probability for particular group for each voxel;
- `Argmax` – is the torch.argmax function which outputs the voxels with the maximum in the final channel, which is a mixed group;
- `Pruning` – is the function that removes all the voxels not chosen by the mask from the input tensor x;
- `Reconstruct` – is the function that reconstructs the dense 3D volume from voxels on the current Stage and all previously memorized voxels from the previous Stages.

**Preprocessing:**
$lr \leftarrow$ uniformly sample a cube of size $32^3$ voxels from $LR$ and concatenate it along the phase channel with a uniformly sampled cube of noise $z$ of size $32^3$ voxels.

**Algorithm 1:** 16x 3D Octree-Based Progressive Growing Generator

**Result:** Process input Low-Resolution 3D segmented micro-CT image and reconstruct 16x Super-Resolution 3D segmented output

```
 1  for Stage = 1 do
 2  |   x ← standard conv block(lr);                              // Standard convolutions on input
 3  |   x ← Minkowski Engine format(x);                           // Convert to ME format
 4  |   x ← Minkowski Engine conv(x);                             // Apply ME convolutions
 5  |   class ← Softmax(Classification(x));                       // Voxel-wise classification
 6  |   mask ← Argmax(class);                                     // Find mixed-group voxels
 7  |   x ← Pruning(x, mask);                                     // Select mixed voxels
 8  |   memorized_Stage ← Pruning(x, ∼ mask);                     // Memorize dense voxels
 9  |   if Stage == 0 then
10  |   |   3D Dense_Stage ← Reconstruct(x, memorized_Stage);     // Reconstruct full volume
11  |   |   return 3D Dense_Stage
12  |   else
13  |   |   return x, memorized_1

14  for Stage = 2 to X do
15  |   x ← Minkowski Engine conv trans(x);                       // Upsample using ME transposed convs
16  |   class ← Softmax(Classification(x));
17  |   mask ← Argmax(class);
18  |   x ← Pruning(x, mask);
19  |   memorized_Stage ← Pruning(x, ∼ mask);
20  |   if Stage == O then
21  |   |   3D Dense_Stage ← Reconstruct(x, memorized_1, ..., memorized_Stage);
22  |   |   return 3D Dense_Stage
23  |   else
24  |   |   return x, memorized_Stage
```

**Appendix E:** 16x Octree-Based Progressive Growing (PG) Super-Resolution (SR) training algorithm for isotropic rocks

**Input:**

- $PG\ G_{Stage\ X}$ – 3D Progressive Growing Generator function for Stage X;
- $PG\ D_{Stage\ X}$ – 2D Progressive Growing Discriminator function for Stage X;
- $LR$ – 3D Low-Resolution training dataset;
- $HR_{Stage\ X}$ – 2D High-Resolution training dataset for Stage X;
- $c$ – coefficient of the voxel-wise loss multiplication;
- $\sigma$ – nearest neighbor downsampling function;
- $N_{Stage\ X}$ – number of epochs per Stage X;
- $n$ – number of iterations per epoch (same for all Stages).

**Preprocessing:**
$HR_{Stage\ X} \leftarrow$ Apply all eight combinations of mirroring and 90° rotations for data augmentation.

---

**Algorithm 2:** 16x Octree-Based Progressive Growing Super-Resolution Training

**Result:** Train PG Generator and Discriminator models for all Stages.

1. **for** *Stage* $X = 1$ **to** 5 **do**
2.    **for** $i = 1$ **to** $N_{Stage\ X}$ **do**
3.       **for** $j = 1$ **to** $n$ **do**
4.          $lr \leftarrow$ Uniformly sample $32^3$ cube from $LR$;
5.          $z \leftarrow$ Sample noise cube of size $32^3$;
6.          $lr \leftarrow$ concatenate$(lr, z)$ along phase channel;
7.          $sr \leftarrow PG\ G_{Stage\ X}(lr);$  // Generate SR volume of size $(32 \cdot 2^{X-1})^3$
8.          $sr_{\text{slices}} \leftarrow$ slice$(sr)$ in xy-, yz-, and xz-planes;
9.          **for** $k = 1$ **to** 3 **do**
10.             $hr \leftarrow$ Sample 128 2D squares of size $(32 \cdot 2^{X-1})^2$ from $HR_{Stage\ X}$;
11.             $l_{gp} \leftarrow$ GP$(sr_{\text{slices plane } k},\ hr,\ PG\ D_{StageX});$  // Gradient Penalty
12.             $l_D \leftarrow PG\ D_{StageX}(sr_{\text{slices plane } k}) - PG\ D_{StageX}(hr) + l_{gp};$
13.             Backpropagate and update $PG\ D_{Stage\ X}$ using $l_D$;
14.          $l_{vw} \leftarrow$ MSE$(lr_{\text{pore space}},\ \sigma(sr_{\text{pore space}}));$  // Voxel-wise loss
15.          $l_G \leftarrow 0;$  // Initialize Generator loss
16.          **for** $k = 1$ **to** 3 **do**
17.             $l_{G\ \text{plane } k} \leftarrow -PG\ D_{Stage\ X}(sr_{\text{slices plane } k}) + c \cdot l_{vw};$
18.             $l_G \leftarrow l_G + l_{G\ \text{plane } k};$
19.          Backpropagate and update $PG\ G_{Stage\ X}$ using $l_G$;

---

**Output:** Trained PG G model for all Stages.